\pdfoutput=1
\documentclass[12pt,a4paper]{article}

\usepackage{ifthen} 
\newboolean{pdflatex}
\setboolean{pdflatex}{true} 

\newboolean{articletitles}
\setboolean{articletitles}{true} 

\newboolean{uprightparticles}
\setboolean{uprightparticles}{false} 

\def\paperauthors{LHCb collaboration} 
\def\paperasciititle{Search for the decays Bds to jpsi gamma at LHCb} 
\def\papertitle{Search for the decays $\Bds\to\jpsi\g$ at \lhcb} 
\def\paperkeywords{{High Energy Physics}, {LHCb}} 
\def\papercopyright{\the\year\ CERN for the benefit of the LHCb collaboration} 
\def\paperlicence{CC BY 4.0 licence}
\def\paperlicenceurl{https://creativecommons.org/licenses/by/4.0/}

\newif\ifEnableSectionTOCLinks
\EnableSectionTOCLinksfalse 

\usepackage[top=1in, bottom=1.25in, left=1in, right=1in]{geometry}

\usepackage{microtype}
\usepackage{lineno}  
\usepackage{xspace} 
\usepackage{caption} 

\usepackage{graphicx}  
\usepackage{color}
\usepackage{colortbl}
\graphicspath{{./figs/}} 

\usepackage{amsmath} 
\usepackage{amssymb}
\usepackage{amsfonts}
\usepackage{upgreek} 

\newcommand*\patchAmsMathEnvironmentForLineno[1]{%
\expandafter\let\csname old#1\expandafter\endcsname\csname #1\endcsname
\expandafter\let\csname oldend#1\expandafter\endcsname\csname
end#1\endcsname
 \renewenvironment{#1}%
   {\linenomath\csname old#1\endcsname}%
   {\csname oldend#1\endcsname\endlinenomath}%
}
\newcommand*\patchBothAmsMathEnvironmentsForLineno[1]{%
  \patchAmsMathEnvironmentForLineno{#1}%
  \patchAmsMathEnvironmentForLineno{#1*}%
}
\AtBeginDocument{%
\patchBothAmsMathEnvironmentsForLineno{equation}%
\patchBothAmsMathEnvironmentsForLineno{align}%
\patchBothAmsMathEnvironmentsForLineno{flalign}%
\patchBothAmsMathEnvironmentsForLineno{alignat}%
\patchBothAmsMathEnvironmentsForLineno{gather}%
\patchBothAmsMathEnvironmentsForLineno{multline}%
\patchBothAmsMathEnvironmentsForLineno{eqnarray}%
}

\usepackage[pdftex,
            pdfauthor={\paperauthors},
            pdftitle={\paperasciititle},
            pdfkeywords={\paperkeywords}]{hyperref}
\usepackage{hyperxmp}
\hypersetup{
    pdfcopyright={Copyright (C) \papercopyright},
    pdflicenseurl={\paperlicenceurl}
}

\usepackage[colorinlistoftodos,textsize=scriptsize]{todonotes}

\usepackage[bottom,flushmargin,hang,multiple]{footmisc}

\usepackage[all]{hypcap} 

\usepackage{xspace} 
\usepackage{upgreek}

\def\lhcb   {\mbox{LHCb}\xspace}

\def\babar  {\mbox{BaBar}\xspace}

\def\belletwo {\mbox{Belle~II}\xspace}

\def\MagUp {\mbox{\em Mag\kern -0.05em Up}\xspace}

\ifthenelse{\boolean{uprightparticles}}%
{
 
 \def\Pgamma      {\ensuremath{\upgamma}\xspace}

 \def\Peta        {\ensuremath{\upeta}\xspace}

 \def\Pmu         {\ensuremath{\upmu}\xspace}

 \def\Ppi         {\ensuremath{\uppi}\xspace}                 
                  
 \def\Prho        {\ensuremath{\uprho}\xspace}

 \def\Ppsi        {\ensuremath{\uppsi}\xspace}

 \def\PDelta      {\ensuremath{\Delta}\xspace}                 
 \def\PXi         {\ensuremath{\Xi}\xspace}                 
 \def\PLambda     {\ensuremath{\Lambda}\xspace}                 
 \def\PSigma      {\ensuremath{\Sigma}\xspace}                 
 \def\POmega      {\ensuremath{\Omega}\xspace}                 
 \def\PUpsilon    {\ensuremath{\Upsilon}\xspace}
 \let\oldPi\Pi
 \def\PPi         {\ensuremath{\oldPi}\xspace}

 \def\PB      {\ensuremath{\mathrm{B}}\xspace}                 
 \def\PD      {\ensuremath{\mathrm{D}}\xspace}                 
 \def\PJ      {\ensuremath{\mathrm{J}}\xspace}                 
 \def\PK      {\ensuremath{\mathrm{K}}\xspace}                 
 \def\PW      {\ensuremath{\mathrm{W}}\xspace}                 
 \def\Pb      {\ensuremath{\mathrm{b}}\xspace}                 
                  
 \def\Pd      {\ensuremath{\mathrm{d}}\xspace}                 
 \def\Pe      {\ensuremath{\mathrm{e}}\xspace}                 
 \def\Pp      {\ensuremath{\mathrm{p}}\xspace}                 

 \def\Ps      {\ensuremath{\mathrm{s}}\xspace}

 \def\thebaroffset{0.0em}
}
{
 
 \def\Pgamma      {\ensuremath{\gamma}\xspace}

 \def\Peta        {\ensuremath{\eta}\xspace}

 \def\Pmu         {\ensuremath{\mu}\xspace}

 \def\Ppi         {\ensuremath{\pi}\xspace}                 
                  
 \def\Prho        {\ensuremath{\rho}\xspace}

 \def\Ppsi        {\ensuremath{\psi}\xspace}                 
                  
 \mathchardef\PDelta="7101
 \mathchardef\PXi="7104
 \mathchardef\PLambda="7103
 \mathchardef\PSigma="7106
 \mathchardef\POmega="710A
 \mathchardef\PUpsilon="7107
 \mathchardef\PPi="7105
 \def\PB      {\ensuremath{B}\xspace}                 
 \def\PD      {\ensuremath{D}\xspace}                 
 \def\PJ      {\ensuremath{J}\xspace}                 
 \def\PK      {\ensuremath{K}\xspace}                 
 \def\PW      {\ensuremath{W}\xspace}                 
 \def\Pb      {\ensuremath{b}\xspace}                 
                  
 \def\Pd      {\ensuremath{d}\xspace}                 
 \def\Pe      {\ensuremath{e}\xspace}                 
 \def\Pp      {\ensuremath{p}\xspace}                 

 \def\Ps      {\ensuremath{s}\xspace}

 \def\thebaroffset{0.18em}
}
\newcommand{\offsetoverline}[2][\thebaroffset]{\kern #1\overline{\kern -#1 #2}}%

\makeatletter
\ifcase \@ptsize \relax
  \newcommand{\miniscule}{\@setfontsize\miniscule{4}{5}}
\or
  \newcommand{\miniscule}{\@setfontsize\miniscule{5}{6}}
\or
  \newcommand{\miniscule}{\@setfontsize\miniscule{5}{6}}
\fi
\makeatother

\DeclareRobustCommand{\optbar}[1]{\shortstack{{\miniscule (\rule[.5ex]{1.25em}{.18mm})}
  \\ [-.7ex] $#1$}}

\def\electron   {{\ensuremath{\Pe}}\xspace}
\def\en         {{\ensuremath{\Pe^-}}\xspace}   
\def\ep         {{\ensuremath{\Pe^+}}\xspace}

\def\mumu       {{\ensuremath{\Pmu^+\Pmu^-}}\xspace}

\def\g      {{\ensuremath{\Pgamma}}\xspace}

\def\W      {{\ensuremath{\PW}}\xspace}

\def\dquark    {{\ensuremath{\Pd}}\xspace}

\def\squark    {{\ensuremath{\Ps}}\xspace}

\def\bquark    {{\ensuremath{\Pb}}\xspace}
\def\bquarkbar {{\ensuremath{\overline \bquark}}\xspace}

\def\pion   {{\ensuremath{\Ppi}}\xspace}
\def\piz    {{\ensuremath{\pion^0}}\xspace}

\def\rhomeson {{\ensuremath{\Prho}}\xspace}

\def\rhop     {{\ensuremath{\rhomeson^+}}\xspace}

\def\kaon    {{\ensuremath{\PK}}\xspace}

\def\KorKbar {\kern \thebaroffset\optbar{\kern -\thebaroffset \PK}{}\xspace}

\def\KS      {{\ensuremath{\kaon^0_{\mathrm{S}}}}\xspace}

\def\Kstarz  {{\ensuremath{\kaon^{*0}}}\xspace}

\def\Kstar   {{\ensuremath{\kaon^*}}\xspace}

\def\Kstarp  {{\ensuremath{\kaon^{*+}}}\xspace}

\newcommand{\etaz}{\ensuremath{\Peta}\xspace}

\def\D       {{\ensuremath{\PD}}\xspace}

\def\DorDbar {\kern \thebaroffset\optbar{\kern -\thebaroffset \PD}\xspace}

\def\Dp      {{\ensuremath{\D^+}}\xspace}
\def\Dm      {{\ensuremath{\D^-}}\xspace}

\def\DpDm    {\ensuremath{\Dp {\kern -0.16em \Dm}}\xspace}

\def\B       {{\ensuremath{\PB}}\xspace}

\def\BorBbar {\kern \thebaroffset\optbar{\kern -\thebaroffset \PB}\xspace}
\def\Bz      {{\ensuremath{\B^0}}\xspace}

\def\Bd      {{\ensuremath{\B^0}}\xspace}

\def\BdorBdbar {\kern \thebaroffset\optbar{\kern -\thebaroffset \Bd}\xspace}
\def\Bu      {{\ensuremath{\B^+}}\xspace}

\def\Bp      {{\ensuremath{\Bu}}\xspace}

\def\Bs      {{\ensuremath{\B^0_\squark}}\xspace}

\def\BsorBsbar {\kern \thebaroffset\optbar{\kern -\thebaroffset \Bs}\xspace}

\def\Bds     {{\ensuremath{\B_{(\squark)}^0}}\xspace}

\def\jpsi     {{\ensuremath{{\PJ\mskip -3mu/\mskip -2mu\Ppsi}}}\xspace}

\def\Y#1S{\ensuremath{\PUpsilon{(#1S)}}\xspace}

\def\proton      {{\ensuremath{\Pp}}\xspace}

\def\LorLbar     {\kern \thebaroffset\optbar{\kern -\thebaroffset \PLambda}\xspace}

\def\BF         {{\ensuremath{\mathcal{B}}}\xspace}
\def\BR         {\BF}

\newcommand{\decay}[2]{\ensuremath{\mathinner{#1\!\to #2}}\xspace}

\def\to                 {\ensuremath{\rightarrow}\xspace}

\def\CP                {{\ensuremath{C\!P}}\xspace}

\def\AT#1     {\ensuremath{A_{\mathrm{T}}^{#1}}\xspace}           

\def\C#1      {\ensuremath{\mathcal{C}_{#1}}\xspace}                       
\def\Cp#1     {\ensuremath{\mathcal{C}_{#1}^{'}}\xspace}                    
\def\Ceff#1   {\ensuremath{\mathcal{C}_{#1}^{\mathrm{(eff)}}}\xspace}        
\def\Cpeff#1  {\ensuremath{\mathcal{C}_{#1}^{'\mathrm{(eff)}}}\xspace}       
\def\Ope#1    {\ensuremath{\mathcal{O}_{#1}}\xspace}                       
\def\Opep#1   {\ensuremath{\mathcal{O}_{#1}^{'}}\xspace}                    

\newcommand{\nospaceunit}[1]{\ensuremath{\text{#1}}}       
\newcommand{\aunit}[1]{\ensuremath{\text{\,#1}}}       

\newcommand{\tev}{\aunit{Te\kern -0.1em V}\xspace}
\newcommand{\gev}{\aunit{Ge\kern -0.1em V}\xspace}
\newcommand{\mev}{\aunit{Me\kern -0.1em V}\xspace}
\newcommand{\kev}{\aunit{ke\kern -0.1em V}\xspace}
\newcommand{\ev}{\aunit{e\kern -0.1em V}\xspace}
 
\newcommand{\mevc}{\ensuremath{\aunit{Me\kern -0.1em V\!/}c}\xspace}
\newcommand{\gevc}{\ensuremath{\aunit{Ge\kern -0.1em V\!/}c}\xspace}
\newcommand{\mevcc}{\ensuremath{\aunit{Me\kern -0.1em V\!/}c^2}\xspace}
\newcommand{\gevcc}{\ensuremath{\aunit{Ge\kern -0.1em V\!/}c^2}\xspace}

\def\mum  {\ensuremath{\,\upmu\nospaceunit{m}}\xspace}

\def\fb   {\ensuremath{\aunit{fb}}\xspace}
\def\invfb   {\ensuremath{\fb^{-1}}\xspace}

\def\ps   {\ensuremath{\aunit{ps}}\xspace}
\def\fs   {\aunit{fs}}

\def\gsim{{~\raise.15em\hbox{$>$}\kern-.85em
          \lower.35em\hbox{$\sim$}~}\xspace}
\def\lsim{{~\raise.15em\hbox{$<$}\kern-.85em
          \lower.35em\hbox{$\sim$}~}\xspace}

\def\pt         {\ensuremath{p_{\mathrm{T}}}\xspace}

\def\ptot       {\ensuremath{p}\xspace}

\def\evtgen     {\mbox{\textsc{EvtGen}}\xspace}

\def\geant      {\mbox{\textsc{Geant4}}\xspace}

\def\photos     {\mbox{\textsc{Photos}}\xspace}

\def\pythia     {\mbox{\textsc{Pythia}}\xspace}

\def\tell1  {TELL1\xspace}
\def\ukl1   {UKL1\xspace}

\newcommand{\eg}{\mbox{\itshape e.g.}\xspace}
\newcommand{\ie}{\mbox{\itshape i.e.}\xspace}

\newcommand{\lhcborcid}[1]{\href{https://orcid.org/#1}{\hspace*{0.1em}\raisebox{-0.45ex}{\includegraphics[width=1em]{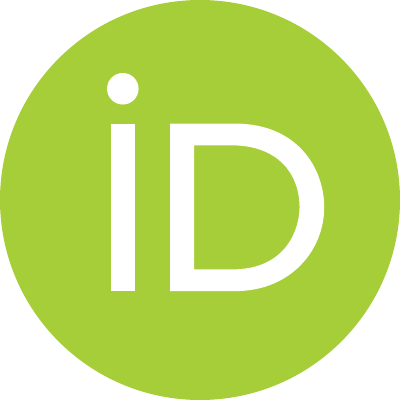}}}}

\hypersetup{
  colorlinks   = true, 
  urlcolor     = blue, 
  linkcolor    = blue, 
  citecolor    = red   
}

\ifEnableSectionTOCLinks
    \usepackage[explicit]{titlesec} 
    
    \let\oldcontentsline\contentsline
    \renewcommand

    \titleformat{\section}{\normalfont\Large\bf}{\hyperlink{tocsection.\thesection}{{\thesection} \parbox[t]{\dimexpr\textwidth-1pc}{#1}}}{1pc}{}

    \titleformat{\subsection}{\normalfont\bf}{\hyperlink{tocsubsection.\thesubsection}{{\thesubsection} \parbox[t]{\dimexpr\textwidth-1pc}{#1}}}{1pc}{}

    \titleformat{name=\section,numberless}[display]{}{}{0pt}{\normalfont\Huge\bfseries #1}
\fi

\usepackage{cite} 
\usepackage{LHCb/mciteplus}

\usepackage{longtable} 
\usepackage{main_def} 
\usepackage{multirow}
\usepackage{siunitx} 

\begin{document}

\renewcommand{\thefootnote}{\fnsymbol{footnote}}
\setcounter{footnote}{1}


\begin{titlepage}
\pagenumbering{roman}

\vspace*{-1.5cm}
\centerline{\large EUROPEAN ORGANIZATION FOR NUCLEAR RESEARCH (CERN)}
\vspace*{1.5cm}
\noindent
\begin{tabular*}{\linewidth}{lc@{\extracolsep{\fill}}r@{\extracolsep{0pt}}}
\ifthenelse{\boolean{pdflatex}}
{\vspace*{-1.5cm}\mbox{\!\!\!\includegraphics[width=.14\textwidth]{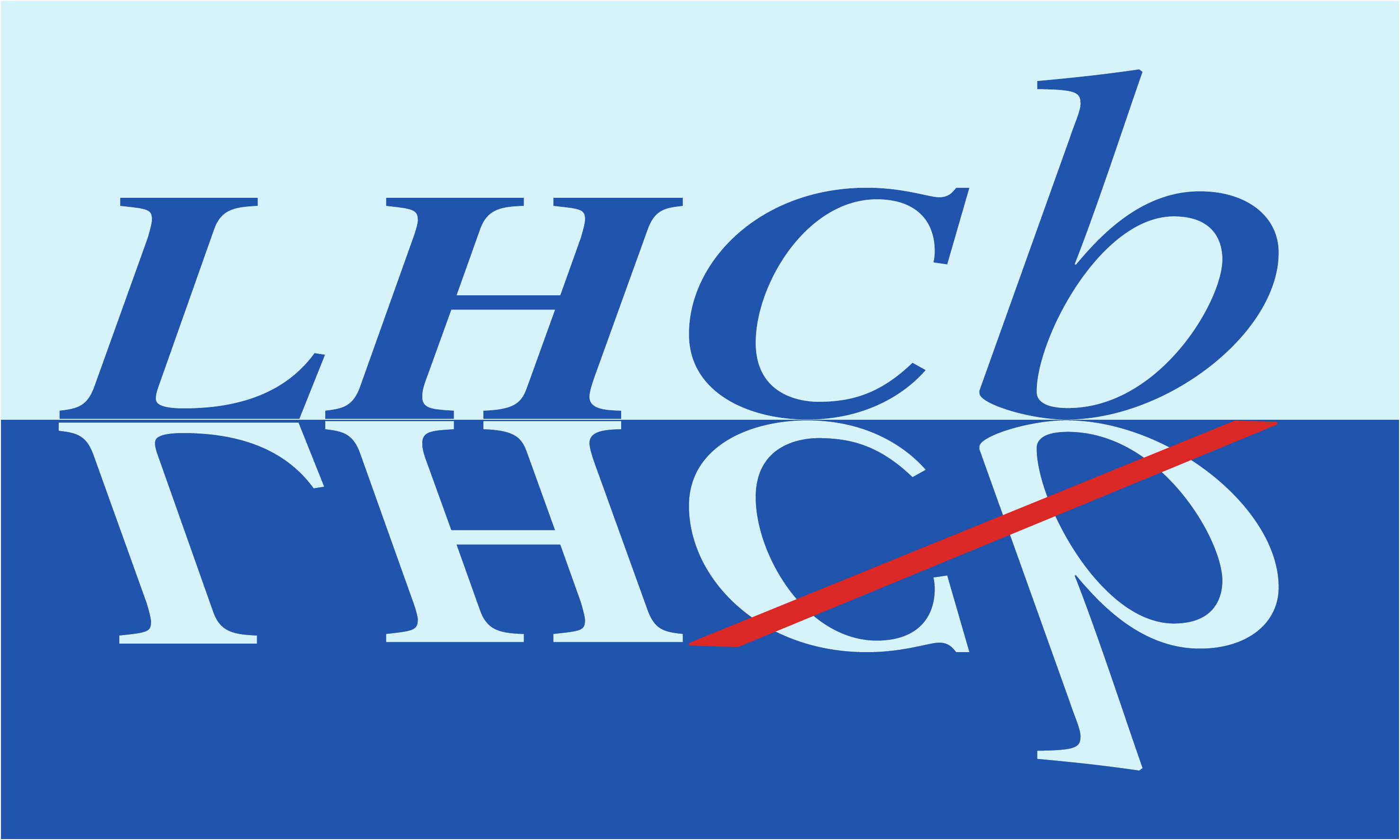}} & &}%
{\vspace*{-1.2cm}\mbox{\!\!\!\includegraphics[width=.12\textwidth]{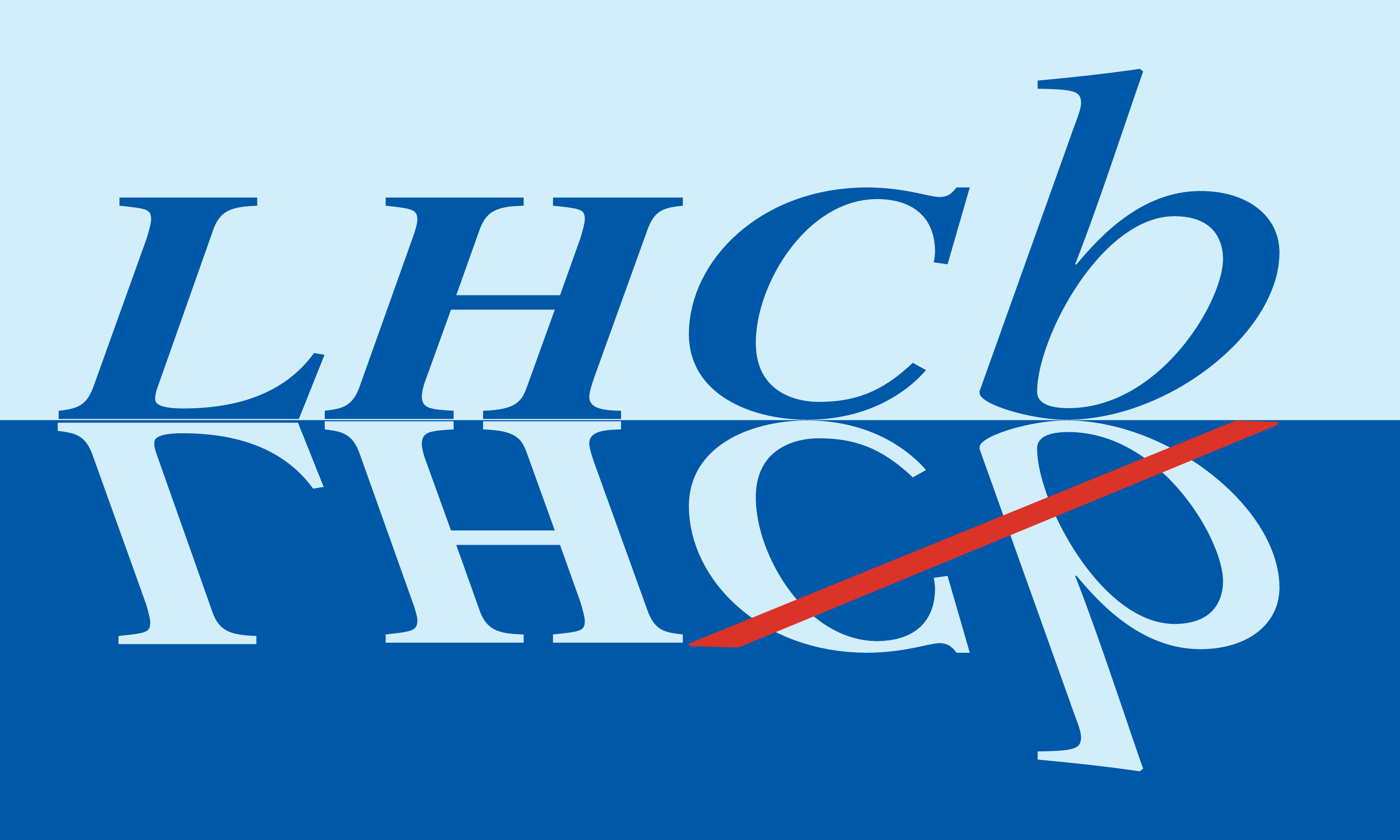}} & &}%
\\
 & & CERN-EP-2026-060 \\  
 & & LHCb-PAPER-2025-065 \\  
 & & April 1, 2026 \\ 
 & & \\
\end{tabular*}

\vspace*{4.0cm}

{\normalfont\bfseries\boldmath\huge
\begin{center}
  \papertitle 
\end{center}
}

\vspace*{2.0cm}

\begin{center}
\paperauthors\footnote{Authors are listed at the end of this paper.}
\end{center}

\vspace{\fill}

\begin{abstract}
  \noindent
  A search for the rare decays \bdstojg is performed with proton-proton collision data collected by the \lhcb experiment, corresponding to integrated luminosities of 3\invfb at centre-of-mass energies of 7 and 8\tev, and 6\invfb at 13\tev.
Assuming no contribution from \btojg decay, an upper limit is set on the branching fraction $\BR(\bstojg) <3.7\times10^{-6}$
at the 90\% confidence level. If instead no contribution from \bstojg decay is assumed, the
limit is $\BR(\btojg)<3.3\times10^{-6}$ at the 90\% confidence level.
These results supersede the previous LHCb results, with the limit for \bstojg improved by a factor of 2.

\end{abstract}

\vspace*{2.0cm}

\begin{center}
  Submitted to
  JHEP
\end{center}

\vspace{\fill}

{\footnotesize 
\centerline{\copyright~\papercopyright. \href{\paperlicenceurl}{\paperlicence}.}}
\vspace*{2mm}

\end{titlepage}


\newpage
\setcounter{page}{2}
\mbox{~}
%
%
%
%


\renewcommand{\thefootnote}{\arabic{footnote}}
\setcounter{footnote}{0}


\cleardoublepage


\pagestyle{plain} 
\setcounter{page}{1}
\pagenumbering{arabic}


\section{Introduction}
\label{sec:Introduction}
\begin{figure}[!b]
  \centering
  \includegraphics[width=0.4\linewidth]{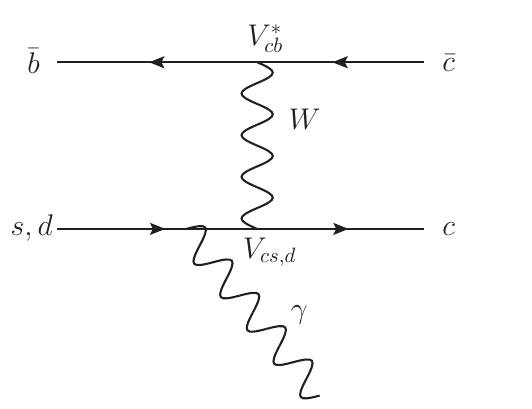}
  \caption{Feynman diagram for the leading contribution to the $\bdstojg$ amplitudes. The photon could also be radiated from the other quarks, 
    but the radiation from heavy quarks is suppressed. Reproduced from Ref.~\cite{LHCb-PAPER-2015-044}.
  }
  \label{fig_feynman}
\end{figure}

The pure annihilation-type radiative decays $\bstojg$ and $\btojg$ have not been observed yet.\footnote{The inclusion of charge-conjugate processes is implied throughout.}
In the Standard Model (SM), these decays proceed via \W boson exchange between the \bquarkbar and the light-flavour quark, as illustrated in Fig.~\ref{fig_feynman}.
Theoretical predictions for the branching fractions in the SM, using a factorization framework that splits the process into a perturbatively calculable short-distance term and a nonperturbative long-distance contribution, are in the range from $1.5\times10^{-7}$ to $5\times10^{-6}$ for $\bstojg$, and at least one order of magnitude lower for the Cabibbo-suppressed decay $\btojg$~\cite{Lu:2003ix,Li:2006xe,Kozachuk:2015kos,Geng:2015ifb}.
The differences in these predictions arise from the choice of the factorisation method and the underlying assumptions.
Therefore, comparing these predictions with an experimental measurement would help test these factorisation schemes.
On the other hand, these decays could be significantly enhanced by the presence of intrinsic charm in \B mesons~\cite{Brodsky:2001yt} or physics beyond the SM (\eg, a right-handed current~\cite{Lu:2003ix}), making them powerful probes of these new physics phenomena.
In addition, measurements of the photon polarisation and charge-parity (\CP) asymmetry, upon observation, would unveil more about the underlying physics of these decays.
Currently, only upper limits on the branching fractions have been reported in experimental searches performed by \babar and \lhcb collaborations~\cite{BaBar:2004lch,LHCb-PAPER-2015-044}.
The most stringent limits at the 90\% confidence level (CL) are $7.3\times10^{-6}$ for the \bstojg and $1.5\times10^{-6}$ for the \btojg decays~\cite{LHCb-PAPER-2015-044}.
The limit for \bstojg is already close to the upper bound of the theoretical predictions.

As an update to the previous \lhcb analysis~\cite{LHCb-PAPER-2015-044}, this paper reports a search using proton-proton (\pp) collision data recorded by the \lhcb experiment, corresponding to integrated luminosities of 3\invfb at centre-of-mass energies of 7 and 8\tev, and 6\invfb at 13\tev, referred to as the Run~1 and Run~2 data samples, respectively. 
The \bdstojg candidates are reconstructed from \decay{\jpsi}{\mumu} decays and photons from their conversion into electron-positron pairs in the detector.
The analysis is optimised for the \bstojg decay, which, unless otherwise specified, is referred to as signal hereafter. To avoid
experimenter's bias, the invariant-mass region around the \Bs mass (5250--5450\mev) was not examined until the full analysis procedure had been finalised.\footnote{Natural units with $\hbar=c=1$ are used throughout.}
The selected dataset is also analysed to search for \btojg decays. The partially reconstructed \B hadron decays, \btojnpi, \btojeta and \bstojeta,
with a photon missing in $\piz\to\g\g$ and $\etaz\to\g\g$ decays, are the main physics backgrounds.

\section{Detector and simulation}
\label{sec:det_sim}
The LHCb detector~\cite{LHCb-DP-2008-001,LHCb-DP-2014-002} is a single-arm forward spectrometer covering the pseudorapidity range $2<\eta<5$.\footnote{The LHCb coordinate system is right-handed, with the $z$ axis pointing along the
beam axis, $y$ the vertical and $x$ the horizontal direction. The $(x, z)$ plane is the bending plane
of the dipole magnet.}
The detector used for this analysis includes a high-precision tracking system consisting of a silicon-strip vertex detector surrounding the \pp interaction region~\cite{LHCb-DP-2014-001}, a large-area silicon-strip detector located upstream of a dipole magnet with a bending power of approximately 4\Tm, and three stations of silicon-strip detectors and straw drift tubes~\cite{LHCb-DP-2013-003,LHCb-DP-2017-001} placed downstream of the magnet.
The tracking system provides a measurement of the momentum, \ptot, of charged particles with a relative uncertainty that varies from 0.5\% at low momentum to 1.0\% at 200\gev.
The minimum distance of a track to a primary \pp collision vertex (PV), the impact parameter (IP), is measured with a resolution of \mbox{$(15+29/\pt)\mum$}, where \pt is the component of the momentum transverse to the beam, in\,\gev.
Particle identification of photons, electrons, and hadrons is provided by two ring-imaging 
Cherenkov detectors, an electromagnetic and a hadronic calorimeter.
Muons are identified by a system composed of alternating layers of iron and multiwire proportional chambers~\cite{LHCb-DP-2012-002}.
The online event selection is performed by a trigger~\cite{LHCb-DP-2012-004,LHCb-DP-2019-001}, which consists of a hardware stage followed by a two-level software stage.
Triggered data further undergo a centralised, offline processing step
to deliver physics-analysis-ready data across the entire \lhcb physics programme~\cite{Stripping}.

Simulation is used to optimise selection requirements, determine efficiencies, and to describe the invariant-mass distribution of the signal candidates.
In the simulation, \pp collisions are generated using \pythia~\cite{Sjostrand:2007gs} with a specific LHCb configuration~\cite{LHCb-PROC-2010-056}.
Decays of unstable particles are described by \evtgen~\cite{Lange:2001uf}, in which final-state radiation is generated using \photos~\cite{davidson2015photos}.
The interaction of the generated particles with the detector, and its response, are implemented using the \geant~\cite{Allison:2006ve} toolkit as described in Ref.~\cite{LHCb-PROC-2011-006}.

\section{Event selection}
\label{sec:sel}
At the hardware trigger stage, selected events are required to contain high-\pt muon or dimuon
candidates, based on information from the muon system.
The first stage of the software trigger performs a partial event reconstruction and requires events to have two well-identified oppositely charged muons 
with a combined invariant mass larger than 2.7\gev. The second stage performs a full event reconstruction, and selected events for further processing if they contain a $\jpsi\to\mumu$ candidate with a
decay vertex well separated from all reconstructed PVs.

In the offline selection, both muons are required to have \mbox{$\pt>500\mev$}, good track-fit qualities, 
and an IP with respect to any PV significantly different from zero.
The two muons should form a good-quality decay vertex.
As the partially reconstructed backgrounds stemming from the \btojnpi and \bstojeta processes can have a reconstructed mass close to the \Bs signal region, using converted photons to improve the resolution also significantly reduce these background sources in the \Bs signal region.
Converted photons are reconstructed, following a similar strategy to that described in Ref.~\cite{LHCb-PAPER-2013-028}, by combining pairs of tracks with opposite charge identified as an electron and a positron, requiring them to have associated clusters in the electromagnetic calorimeter and a good track-fit quality.
The energy loss of electrons and positrons by emission of bremsstrahlung photons is recovered by adding the energies of the reconstructed photons consistent with originating from the track.
Due to different mass resolutions, the candidates are separated into two categories, based on where the photon converts in the detector.
Conversions which occur early enough for the converted electrons and positrons to be reconstructed in the vertex detector are referred to as $\textit{long}$ because the tracks pass through the full tracking system, whilst those that convert late enough such that track segments of the electrons and positrons cannot be formed in the vertex detector are referred to as $\textit{downstream}$~\cite{LHCb-DP-2013-002}.
The electron-positron pair candidate is required to have a reconstructed invariant mass smaller than $60\mev$~($100\mev$) for the $\textit{long}$ ($\textit{downstream}$) case and \mbox{$\pt>1\gev$}.
An additional tighter invariant-mass selection is applied to photon candidates whose conversion $z$-coordinate position is close to the PV, to reduce the contamination from the Dalitz decay $\piz\to\ep\en\g$ and combinatorial background.

The \jpsi and \g candidates are combined to form \Bds candidates.
To improve the resolution on
the reconstructed \B-hadron mass, a kinematic fit is performed~\cite{Hulsbergen:2005pu}, in which the dimuon mass is constrained to the known \jpsi mass value~\cite{PDG2024} and the \B hadron to originate from its associated PV, defined as the PV that fits best to the flight direction of the \B-hadron candidate.
The candidates are further required to have an invariant mass in the range \mbox{$4000<\mjpsig<7000\mev$} and \mbox{$\pt>2\gev$}, significant flight distance from the associated PV, and a good consistency between their momentum and flight direction.

Two boosted decision tree (BDT) classifiers~\cite{Breiman,AdaBoost} are used to increase the signal significance.
The first classifier is dedicated to rejecting combinatorial background, where the \jpsi and \g candidates come from different sources, while the second suppresses the partially reconstructed \B-hadron decays.
The input variables are primarily kinematic and geometric, alongside isolation criteria used to reject background containing additional tracks in a cone around the \B-hadron direction, and the fit quality of the decay topology. For the \mylong category, particle identification information of the electrons is also included to reduce background candidates where charged hadrons are misidentified as electrons.
Separate BDTs are trained for the \mylong and \mydownstream categories. The combinatorial background is represented by candidates in the low-mass (\mbox{$4000<\mjpsig<4800\mev$}) and in the high-mass (\mbox{$5900<\mjpsig<6800\mev$}) sideband region of the data, while simulation samples are used for the signal and the partially reconstructed backgrounds \btojnpi, \bdstojeta, \btojrho and \btojkst.\footnote{The symbol \rhop is used to refer to the $\rhomeson(770)^{+}$ meson and \Kstar is used to refer to the $\Kstar(892)$ meson throughout the paper.}
The k-fold cross-validation method~\cite{kFold} with $k=7\,(5)$ for the first (second) BDT is used to make full use of the available statistics.
The requirements on the BDT responses are optimised simultaneously by maximising the Punzi figure of merit \mbox{$\varepsilon_{\text{S}}^{\text{BDT}}/(3/2+\sqrt{N_\text{B}})$}~\cite{Punzi:2003bu}, where $\varepsilon_{\text{S}}^{\text{BDT}}$ is the BDT efficiency for the signal calculated using simulation and $N_{\text{B}}$ the estimated background yield in the signal region extrapolated from a background-only fit to the data in the mass sidebands.

\begin{figure}
    \centering
    \includegraphics[width=0.49\linewidth]{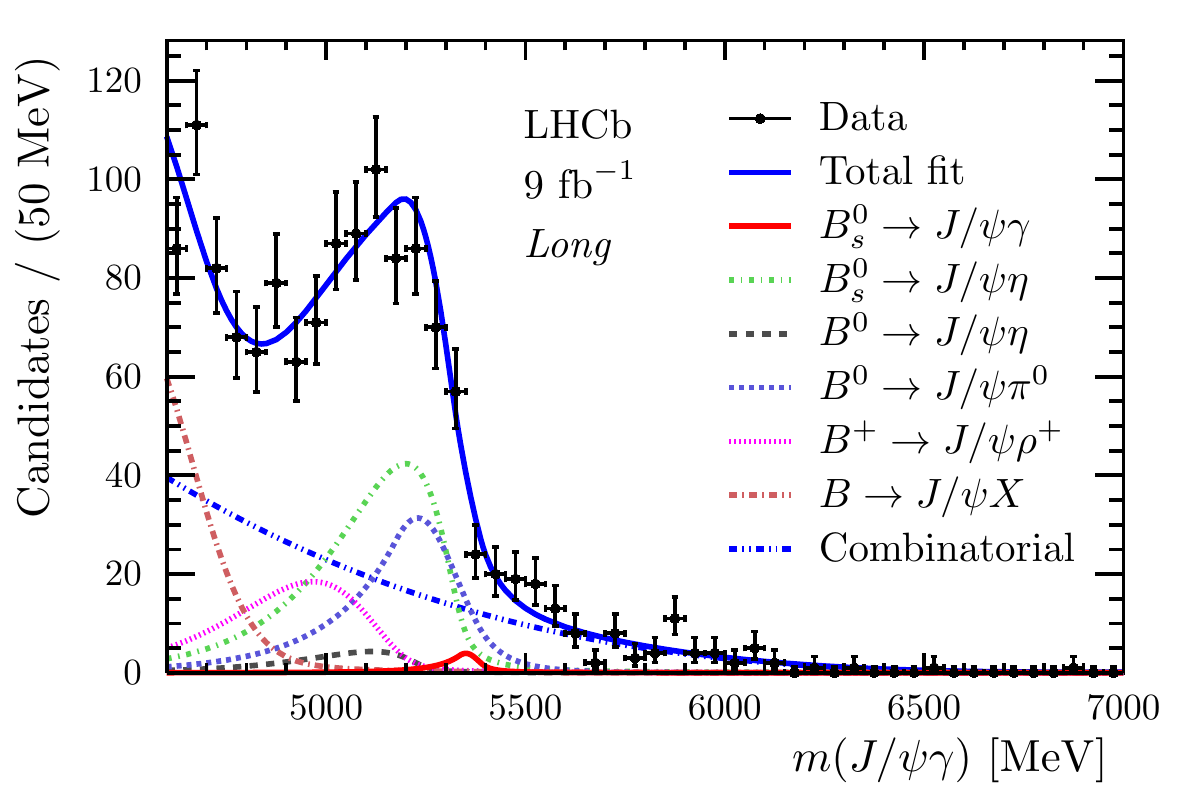}
    \includegraphics[width=0.49\linewidth]{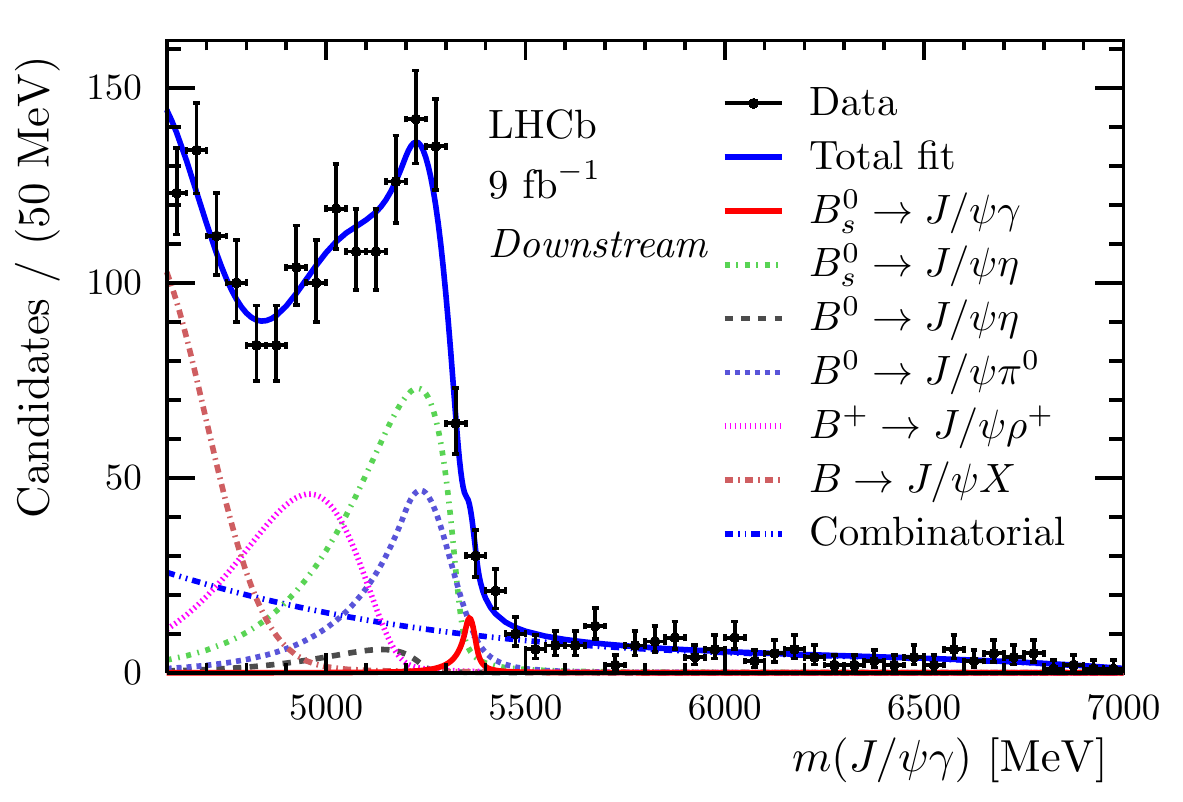}
    \caption{Invariant-mass distributions for the \bstojg decay search in the (left) \mylong and (right) \mydownstream categories, with the result of the fit described in the text also shown.}
    \label{fig:fit}
\end{figure}

Candidates with reconstructed invariant mass \mbox{$4600<\mjpsig<7000\mev$} are used for the search of \bstojg and \btojg decays.
A negligible fraction of the events contains multiple candidates and all are kept.
After the final selection, there are 1351 candidates in the \mylong category and 1842 candidates in the \mydownstream one.
The distributions of the reconstructed invariant mass of the candidates in both categories are shown in Fig.~\ref{fig:fit}.

The selection efficiencies for signal and background decays with the \B hadron in the \mbox{$2<\eta<5$} fiducial region are estimated using simulation, and are listed in Table~\ref{tab:eff_br}.
The \btokstg decay has a similar kinematic topology to the signal process and a large sample size.
Therefore, the \btokstg decay reconstructed with converted photons is used to check the agreement between data and simulation in \Bds kinematics, electron particle identification, and event multiplicity, and to improve them by applying correction factors to the signal and partially reconstructed background simulations.
The fractions of \Bz and \Bs candidates that fall in the fiducial region are assumed to be the same, neglecting the slight \pt-dependence of the ratio of the \Bz and \Bs fragmentation fractions~\cite{LHCb-PAPER-2020-046}.
In general, the background modes have a smaller offline selection efficiency than the signal, primarily because the photon has a lower momentum.
\begin{table}
    \centering
    \caption{Branching fractions used in the fit and total efficiencies in the \mylong and \mydownstream categories for all signal and background decays. 
    The efficiencies are calculated for the \B hadrons in the \mbox{$2<\eta<5$} region.
    The uncertainties on
    the efficiencies are from the simulated sample statistics only.}
    \label{tab:eff_br}
    \begin{tabular}{
        l
        S[table-format=2.2(4)]
        S[table-format=2.4(4)]
        S[table-format=2.4(4)]}
        
        \hline
        \multirow{2}{*}{Process} &  {\multirow{2}{*}{$\BR\ [\times10^{-5}]$}} &  \multicolumn{2}{c}{Efficiency [$\times10^{-4}$]} \\
                &                      &  \mylong  & \mydownstream \\ \hline
        \bstojg     & {--}              & 19.73\pm0.20 & 26.74\pm0.21 \\
        \btojnpi    & 1.73\pm0.07  & 8.09\pm0.11 & 8.91\pm0.10 \\
        \btojeta    & 0.99\pm0.10  & 5.08\pm0.11 & 7.27\pm0.11 \\
        \bstojeta   & 45.8\pm3.4   & 5.31\pm0.11 & 7.48\pm0.12 \\
        \btojrho    & {--}              & 1.41\pm0.04 & 1.36\pm0.03 \\
        \hline
    \end{tabular}
\end{table}

\section{Mass spectra fits}
\label{sec:bf_def}
As neither the \bstojg nor the \btojg decays have been observed to date, the two decays are searched for independently. In each search, an unbinned maximum-likelihood fit to the $\mjpsig$ distribution in the \mylong and \mydownstream categories is performed simultaneously.
In the fit for the \bstojg decay search, the branching fraction for \btojg is set to zero, and when searching for the \btojg decay, 
the branching fraction for \bstojg is set to zero. The fit for the \bstojg decay search is described in detail below, and the \btojg decay
is also searched for following a similar procedure.

The $\mjpsig$ distribution of the signal process is primarily shaped by the invariant-mass resolution, which is dominated by the energy resolution of the photon.
The resolution is modelled using a modified Crystal Ball function~\cite{Skwarnicki:1986xj}~(DSCB), which contains a Gaussian core
and power-law tails on both sides of the peak. Values of the parameters are obtained by fitting the simulated signal sample. The standard deviation of the
Gaussian core is \mbox{$26.5\pm1.4\,(14.6\pm0.4)\mev$} for the \mylong (\mydownstream) category in the simulated sample.
The value is corrected with a scale factor of
\mbox{$1.14\pm0.10\,(1.04\pm0.06)$} to better describe the resolution in data, based on the comparison of \btokstg decays in data and simulation.
The \mydownstream category has a better resolution as electrons in the \mylong category have larger probability to lose energy through bremsstrahlung.

The \bstojeta, \btojeta, \btojrho, and \btojkst decays are described by ARGUS functions~\cite{ARGUS:1990hfq} each convolved with the resolution function.
The shape of the partially reconstructed background from the \btojnpi decay includes some fully reconstructed decays, as the bremsstrahlung correction can recover the missing photon from the \piz decay.
The ARGUS function cannot describe the \btojnpi shape well, so a DSCB function is used instead.
The partially reconstructed decays missing a kaon or multiple hadrons, which are well below the \Bds mass peak, are modelled using the simulated \btojkst sample. Hereafter, this component is referred to as \btojX, where \PX represents other particles that can decay into final states including one or more heavy hadrons and one \piz meson.
Similarly, the \btojks decays are included in the \btojrho description. The values of the shape parameters in these
background models are obtained from the corresponding simulated samples.
Finally, the combinatorial background is modelled using a third-order Bernstein polynomial function with the parameters left free to vary in the fit.

The yields of the signal process and the \btojeta and \bstojeta decays are normalised to the yield of \btojnpi in the \mydownstream category 
according to the branching fraction, with $\mathcal{B}(\eta\to\g\g)=0.394\pm0.002$ also taken into account, and selection efficiency ratios listed in Table~\ref{tab:eff_br}:
\begin{equation}
N_\text{proc}^{\text{cat}}=\frac{f_{i}}{f_{\dquark}}\cdot\frac{\BF_\text{proc}\cdot\varepsilon_\text{proc}^{\text{cat}}}{\BF_{\btojnpi}\cdot\varepsilon^{\mydownstream}_{\btojnpi}}\cdot N_{\btojnpi}^{\mydownstream}, 
    \label{eq:normal}
\end{equation}
where $N$ is the yield, the superscript ``cat'' refers to the \mylong or \mydownstream categories,
the subscript ``proc'' is either \bstojg, \btojeta or \bstojeta, $f_{i}$ is the corresponding \B-hadron fragmentation 
fraction ($i=\dquark, \squark$), \BF and $\varepsilon$ are the branching fraction and efficiency, respectively.
The yield of \btojnpi in \mydownstream category and \sigBr are free parameters in the fit while other parameters are
constrained to the existing measurements.
The value of $\BR(\btojnpi)$ is a combination of the recent \belletwo measurement,
\mbox{$\BR(\btojnpi) = (2.00 \pm 0.12 \pm 0.09) \times 10^{-5}$}~\cite{Belle-II:2024hqw}, and the average of the previous measurements,
\mbox{$\BR(\btojnpi) = (1.65 \pm 0.08) \times 10^{-5}$}~\cite{PDG2024}.
For $\BR(\bstojeta)$ and $\BR(\btojeta)$ the recent measurements from LHCb~\cite{LHCb-PAPER-2025-025} are used.
The ratio of fragmentation fractions $\fs/\fd=0.2502\pm 0.0078$, calculated as the weighted average at three different collision energies~\cite{LHCb-PAPER-2020-046}, is used.
The precision of these external measurements is significantly improved compared to those used in the previous analysis only using Run~1 data~\cite{LHCb-PAPER-2015-044}.
As the yields of these processes in the \mylong and \mydownstream categories are also related by the efficiency ratios and branching fraction ratios, they are not all independent.

The yields of \btojrho and \btojkst decays in both the \mylong and \mydownstream categories are left as free parameters of the fit to allow contributions from \btojks and other \btojX decays.

\begin{figure}
    \centering
    \includegraphics[width=0.7\linewidth]{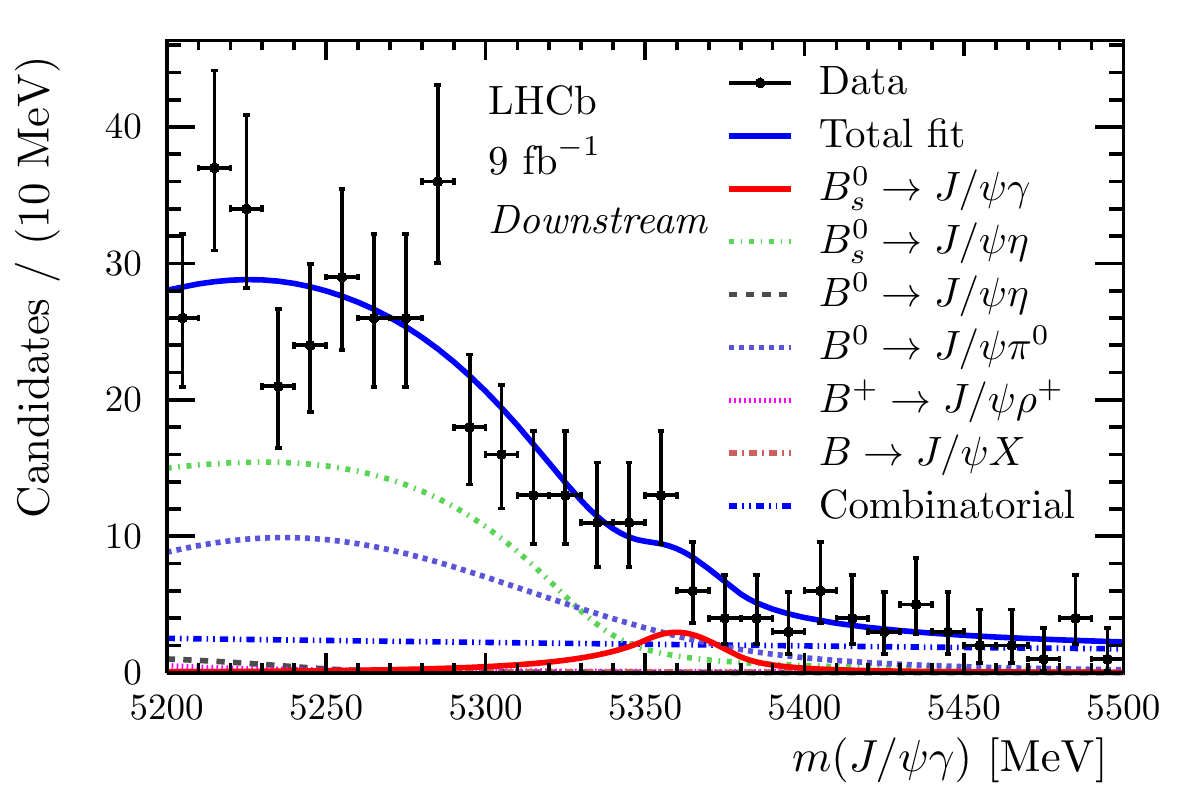}
    \caption{Invariant-mass distribution of the \mydownstream category signal candidates in a narrow mass window around the \Bs mass with the fit result superimposed.}
    \label{fig:narrow_window}
\end{figure}

In the final fit, Gaussian constraints are applied to the nuisance parameters (shape parameters obtained from the fit to simulation, efficiencies, and external measurements), and the correlations are taken into account when applicable.
For example, the shape parameters are constrained using the values and correlation matrices from fits to the simulated sample.
To evaluate the statistical uncertainty in isolation, the values of these nuisance parameters are fixed to their best fit values.
The fit results for the \bstojg decay search are shown in Fig.~\ref{fig:fit}.
The resulting branching fraction is
\begin{equation*}
  \BR(\bstojg) = (1.77\pm0.96)\times10^{-6},
\end{equation*}
where the uncertainty is statistical only.
Thanks to its higher efficiency and better resolution, the sensitivity is driven by the \mydownstream category.
The breakdown of the \mydownstream category yields in various invariant-mass regions is shown in Table~\ref{tab:yield_bkg}.
As expected, the main contribution in the \bstojg region comes from \btojnpi, \bstojeta decays, and combinatorial background.
The \mydownstream category data and fit result in a narrow mass window around the \Bs mass with finer binning are shown in Fig.~\ref{fig:narrow_window}.
No significant signal peak is evident in the data.

\begin{table}[!b]
    \centering
    \caption{Yields for the signal and background processes in the \mydownstream category for the \bstojg decay search.
    The Gaussian constraints are applied to the nuisance parameters.
    The uncertainties are propagated from the free parameters in the fit. The $\sigma$ in the \Bds mass window definition is the standard deviation
  of the Gaussian core of the resolution function.}
    \label{tab:yield_bkg}
    \begin{tabular}{
        l
        S[table-format=3.0(2)]
        S[table-format=2.3(3)]
        S[table-format=3.2(2)]
        }
        \hline
        {Process} & {Mass range} & {$m(\Bs)\pm3\sigma$} & {$m(\Bd)\pm3\sigma$} \\
        \hline
        \bstojg & 17 \pm 10 & 13 \pm 7 & 2.4 \pm 1.4 \\
        \btojnpi & 223 \pm 17 & 21.4 \pm 1.7 & 62 \pm 5 \\
        \bstojeta & 493 \pm 31 & 18.7 \pm 1.2 & 113 \pm 7 \\
        \btojeta & 41 \pm 5 & 0.220 \pm 0.027 & 1.64 \pm 0.20 \\
        \btojrho & 335 \pm 49 & 0.66 \pm 0.10 & 1.35 \pm 0.20 \\
        \btojX & 294 \pm 35 & 0.41 \pm 0.05 & 0.53 \pm 0.06 \\
        Combinatorial & 417 \pm 70 & 17.7 \pm 3.0 & 19.7 \pm 3.3 \\
        \hline
    \end{tabular}
\end{table}

In the mass spectra fit for the \btojg decay search, the \Bz signal shape is the same as the one for \bstojg decays but shifted 
by the known mass difference between \Bs and \Bz mesons of 87.26\mev~\cite{PDG2024}. The selection efficiency is assumed to
be the same as for the \bstojg decay, and the fragmentation fraction is replaced with the corresponding one. The fit result is 
\begin{equation*}
  \BR(\btojg) = (0.77\pm0.62)\times10^{-6},
\end{equation*}
with the statistical uncertainty only considered. The background yields are similar to those given in Table~\ref{tab:yield_bkg}.

\section{Systematic uncertainties}
\label{sec:sys_uncer}
Systematic uncertainties arise from the limited knowledge of the mass spectra of signal and background processes, as well as from other normalisation parameters involved
in the fit.
Their contributions are fully incorporated into the likelihood function, and their values are obtained
by allowing the nuisance parameters to vary within their uncertainties.
Each contribution of systematic uncertainty source is calculated by comparing the uncertainty of the fit results
with and without the relevant parameters fixed, with all other nuisance parameters fixed to their best fit values.
The main systematic uncertainties are summarised in Table~\ref{tab:sys_uncer} and their estimation for the \bstojg decay search is described in more detail below.

\begin{table}[b]
    \centering
    \caption{Relative systematic and statistical uncertainties for the \sigBrs measurement. The systematic uncertainty from the fit accounts for correlations among the nuisance parameters.}
    \begin{tabular}{
        l
        S[table-format=2.1(1)]
        S[table-format=2.1(1)]
        }
    \hline
    Source & $\bstojg$~[\%] & $\btojg$~[\%] \\ \hline
       \quad Mass resolution (core) & 15.4 & 8.8 \\
       \quad Mass resolution (tail)  & 1.3 & 2.0 \\
      \quad Combinatorial background model & 11.8 & 1.6 \\
       \quad Partially reconstructed background model & 8.8 & 24.4 \\
       \quad Selection efficiency & 3.6 & 5.8 \\
       \quad Branching fraction & 8.9 & 5.8 \\
    Systematic (sum in quadrature) & 23.2 & 27.3 \\
         \hline
    Systematic (from fit) & 23.4 & 27.9\\\hline
    Statistical & 54.5 & 80.7 \\\hline
    \end{tabular}
    \label{tab:sys_uncer}
\end{table}

The uncertainty due to the limited knowledge of the mass resolution is assessed for the Gaussian core and the tails separately.
The width of the Gaussian core and the scale factor are also allowed to vary within their uncertainties, 
which are obtained from the simulated signal sample fit and the comparison of \btokstg decays in data and simulation.
The resulting relative systematic contribution is 15.4\%, and, as expected from the small signal yield, it dominates the systematic uncertainty. The tail parameters are allowed to vary as well within their uncertainties
from the simulated signal sample fit, with a resulting small uncertainty.
As the signal shape is just the shifted resolution function, in this way the uncertainty related to the mismodelling of 
the signal decay is also included.

The same method is used for the shapes of all background processes. The variation of the related shape parameters brings 11.8\% uncertainty
from the combinatorial background and 8.8\% from the partially reconstructed backgrounds.

The selection efficiencies enter into the normalisation of the signal and background processes. Three contributions are considered: 
the finite size of the simulated sample, the mismodelling of the \B-hadron kinematics, and the mismodelling of the lifetime of the signal process. While the first contribution is included in the efficiency calculation in Table~\ref{tab:eff_br}, for the second a weighting process is applied to the simulated sample to improve the agreement with the data. The weights are calculated by comparing the kinematics of \btokstg decays in data and simulation. Finally, for the third contribution, since the \CP composition of the signal is unknown, the efficiency is evaluated under two extreme scenarios corresponding to purely \CP-even and \CP-odd states, with lifetimes fixed to $1.429\pm0.006\ps$ and $1.622\pm0.008\ps$~\cite{PDG2024}, respectively. The efficiency is recomputed in each case and compared to the baseline value, and the maximum deviation is assigned as systematic uncertainty.
The total uncertainty on the selection efficiency is included in the mass spectra fit as Gaussian constraints, and amounts to about 3.6\%.

The branching fractions of the \btojnpi, \bstojeta, \btojeta and $\eta\to\g\g$ decays, and the fragmentation fractions are used to relate the yields of 
these processes as per Eq.~\ref{eq:normal}. To reflect their uncertainties, they are also constrained using Gaussian functions in the likelihood fit. The contribution from the branching fractions to the systematic uncertainty amounts to 8.9\%, while the contribution from fragmentation fractions is negligible.

The total relative systematic uncertainty from the mass spectra fit is approximately 23.4\%, which is close to the value of 23.2\% obtained by summing the contributions from each source under the assumption of uncorrelated uncertainties, and is significantly smaller than the statistical uncertainty.

For the \btojg decay, contributions from the various systematic uncertainty sources are also summarized in Table~\ref{tab:sys_uncer}. In contrast to the \bstojg analysis,
the dominant contribution arises from the partially reconstructed background, a result that is expected given that this background category 
dominates in the \Bz signal region.

\section{Results}
\label{sec:result}
\begin{figure}
    \centering
    \includegraphics[width=0.49\linewidth]{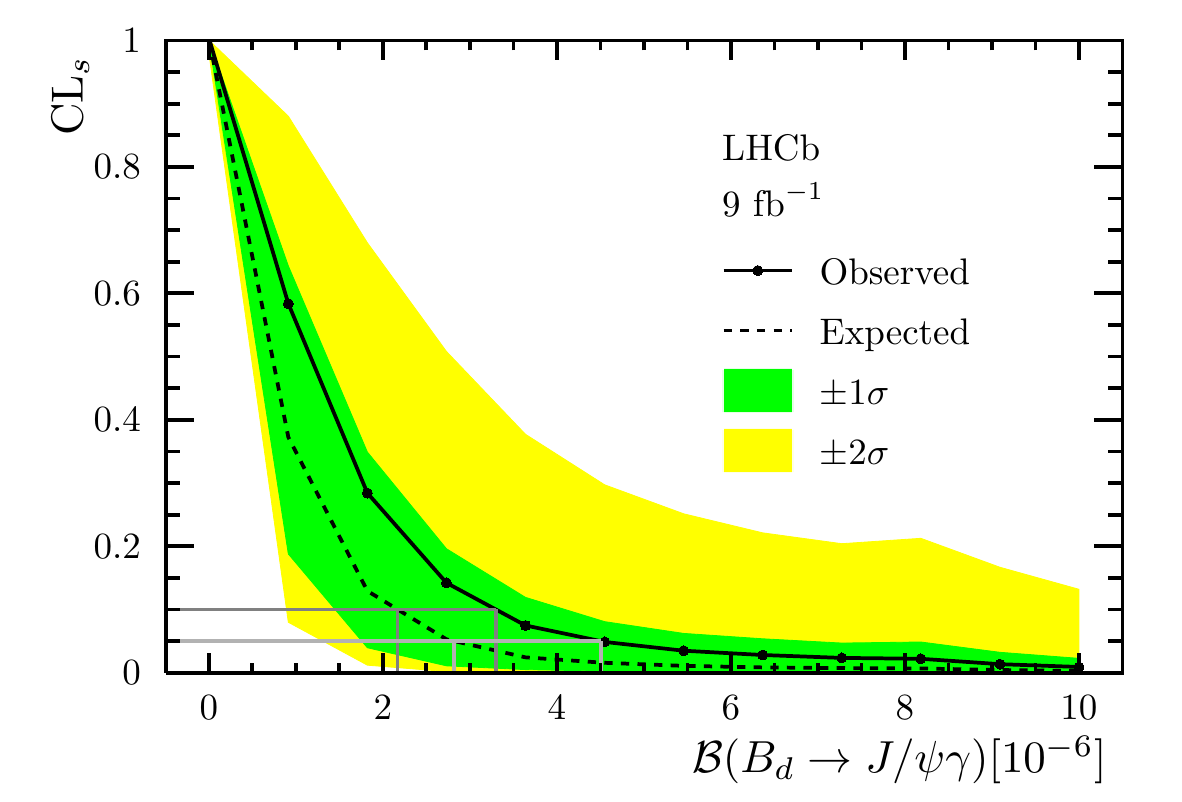}
    \includegraphics[width=0.49\linewidth]{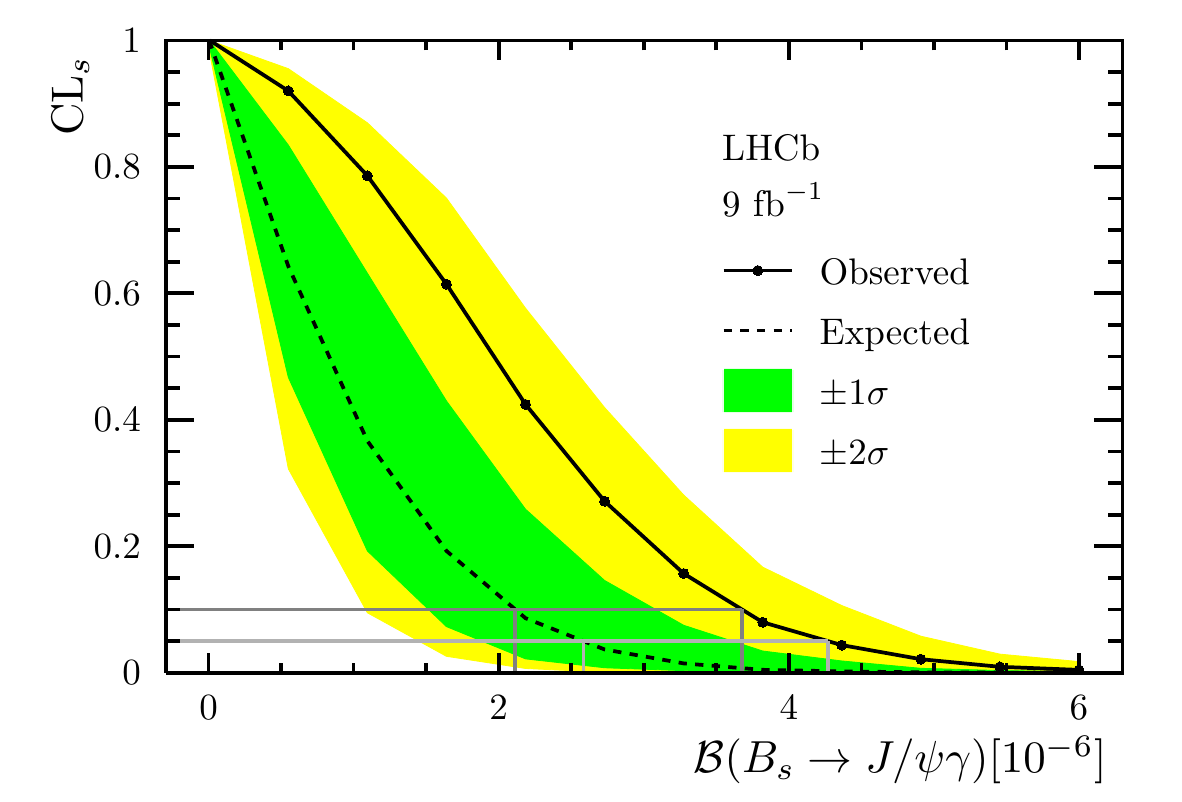}
    \caption{Expected (dashed black line) and observed (black line and points) \CLs as a function of the branching fraction hypothesis for (left) \btojg and (right) \bstojg decays, with (green region) 1$\sigma$ and (yellow region) 2$\sigma$ uncertainty bands. The observed and expected limits are marked by gray horizontal and vertical lines for both 90\% and 95\% CL.}
    \label{fig:cls}
\end{figure}

With the systematic uncertainties included, the result for the \bstojg and \btojg processes are
\begin{equation*}
    \begin{aligned}
        \BR(\bstojg) = (1.77\pm0.96\pm0.41)\times10^{-6},\\
        \BR(\btojg) = (0.77\pm0.62\pm0.21)\times10^{-6},
    \end{aligned}
\end{equation*}
where the first uncertainty is statistical and the second is systematic.
These results are consistent with the background-only hypothesis with a significance below $2\sigma$ for both \Bs and \Bz signal.
Hence, the \CLs method~\cite{CLs,Junk:1999kv} is used to set upper limits on the branching fractions.
The profile likelihood ratio is used as the test statistic for the \CLs calculations.
Pseudoexperiments are generated in order to determine the observed and expected exclusion CL of the branching fraction value.
The observed and expected \CLs exclusions are shown as a function of the hypothesised branching fraction for 
\bstojg and \btojg decays in Fig.~\ref{fig:cls}.
The upper limits of the branching fractions are determined to be
\begin{equation*}
    \begin{aligned}
        \BR(\bstojg)<3.7\,(4.3)\times10^{-6}~\text{at 90\,(95)\% CL},\\
        \BR(\btojg) <3.3\,(4.5)\times10^{-6}~\text{at 90\,(95)\% CL}.
    \end{aligned}
\end{equation*}
The \CLs value for the perturbative QCD prediction \mbox{$\sigBr=5\times10^{-6}$}~\cite{Li:2006xe} is 0.019, \ie it is excluded at the 98.1\% CL.

\section{Conclusion}
\label{sec:conclu}

A search for the decays \btojg and \bstojg is performed with data collected by the \lhcb experiment in Run~1 and Run~2, corresponding to an integrated luminosity of 9\invfb.
The two decay processes are searched for independently: for the \bstojg decay search, no contribution from \btojg decays is assumed, and vice versa for the \btojg decay search.
No significant signal is observed in both channels and upper limits on the branching fractions are set to be \mbox{$\BR(\bstojg)<3.7\,(4.3)\times10^{-6}$} and \mbox{$\BR(\btojg)<3.3\,(4.5)\times10^{-6}$} at 90\,(95)\% CL. These results supersede those of the \lhcb Run~1 analysis~\cite{LHCb-PAPER-2015-044}, with 
the limit for the \bstojg decay improved by a factor of 2. The theoretical prediction
based on perturbative QCD~\cite{Li:2006xe}, $5\times 10^{-6}$, is excluded at the 98.1\% CL. Comparison with the predictions of other models, \eg $1.4\times10^{-6}$ predicted in Ref.~\cite{Lu:2003ix}
and $(7.2\pm0.7)\times10^{-7}$ predicted in Ref.~\cite{Geng:2015ifb}, will be reachable with future larger datasets.
Although the sensitivity of this analysis is optimised for the \bstojg signal, the expected limit for the \btojg decay is also better than the Run~1 analysis. 
However, the observed upper limit is looser than the Run~1 limit. 
As the observed \CLs values are well within the $1\sigma$ band of the background-only hypothesis and the SM prediction of the signal yields for \bstojg and \btojg decays
are negligible with respect to background, this is attributed to statistical fluctuations of the background.

\section*{Acknowledgements}
%
%
\noindent We express our gratitude to our colleagues in the CERN
accelerator departments for the excellent performance of the LHC. We
thank the technical and administrative staff at the LHCb
institutes.
We acknowledge support from CERN and from the national agencies:
ARC (Australia);
CAPES, CNPq, FAPERJ and FINEP (Brazil); 
MOST and NSFC (China); 
CNRS/IN2P3 and CEA (France);  
BMFTR, DFG and MPG (Germany);
INFN (Italy); 
NWO (Netherlands); 
MNiSW and NCN (Poland); 
MEC/IFA (Romania); 
MICIU and AEI (Spain);
SNSF and SER (Switzerland); 
NASU (Ukraine); 
STFC (United Kingdom); 
DOE NP and NSF (USA).
We acknowledge the computing resources that are provided by ARDC (Australia), 
CBPF (Brazil),
CERN, 
IHEP and LZU (China),
IN2P3 (France), 
KIT and DESY (Germany), 
INFN (Italy), 
SURF (Netherlands),
Polish WLCG (Poland),
IFIN-HH (Romania), 
PIC (Spain), CSCS (Switzerland), 
GridPP (United Kingdom),
and NSF (USA).  
We are indebted to the communities behind the multiple open-source
software packages on which we depend.
Individual groups or members have received support from
RTP (Australia), 
Key Research Program of Frontier Sciences of CAS, CAS PIFI, CAS CCEPP (China); 
Minciencias (Colombia);
EPLANET, Marie Sk\l{}odowska-Curie Actions, ERC and NextGenerationEU (European Union);
A*MIDEX, ANR, IPhU and Labex P2IO, and R\'{e}gion Auvergne-Rh\^{o}ne-Alpes (France);
Alexander-von-Humboldt Foundation (Germany);
ICSC (Italy); 
Severo Ochoa and Mar\'ia de Maeztu Units of Excellence, GVA, XuntaGal, GENCAT, InTalent-Inditex and Prog.~Atracci\'on Talento CM (Spain);
the Leverhulme Trust, the Royal Society and UKRI (United Kingdom).





\addcontentsline{toc}{section}{References}
\bibliographystyle{LHCb/LHCb}
\bibliography{main,main_ref,LHCb/standard,LHCb/LHCb-PAPER,LHCb/LHCb-CONF,LHCb/LHCb-DP,LHCb/LHCb-TDR,LHCb/LHCb-PUB}

\clearpage
\centerline
{\large\bf LHCb collaboration}
\begin
{flushleft}
\small
R.~Aaij$^{38}$\lhcborcid{0000-0003-0533-1952},
M. ~Abdelfatah$^{69}$,
A.S.W.~Abdelmotteleb$^{57}$\lhcborcid{0000-0001-7905-0542},
C.~Abellan~Beteta$^{51}$\lhcborcid{0009-0009-0869-6798},
F.~Abudin\'en$^{59}$\lhcborcid{0000-0002-6737-3528},
T.~Ackernley$^{61}$\lhcborcid{0000-0002-5951-3498},
A. A. ~Adefisoye$^{69}$\lhcborcid{0000-0003-2448-1550},
B.~Adeva$^{47}$\lhcborcid{0000-0001-9756-3712},
M.~Adinolfi$^{55}$\lhcborcid{0000-0002-1326-1264},
P.~Adlarson$^{87}$\lhcborcid{0000-0001-6280-3851}
% ...

{\footnotesize \it

% ...
$^{38}$Nikhef National Institute for Subatomic Physics, Amsterdam, Netherlands\\
% ...
$^{47}$Instituto Galego de F{\'\i}sica de Altas Enerx{\'\i}as (IGFAE), Universidade de Santiago de Compostela, Santiago de Compostela, Spain\\
% ...
$^{51}$Physik-Institut, Universit{\"a}t Z{\"u}rich, Z{\"u}rich, Switzerland\\
% ...
$^{55}$H.H. Wills Physics Laboratory, University of Bristol, Bristol, United Kingdom\\
% ...
$^{57}$Department of Physics, University of Warwick, Coventry, United Kingdom\\
% ...
$^{59}$School of Physics and Astronomy, University of Edinburgh, Edinburgh, United Kingdom\\
$^{60}$School of Physics and Astronomy, University of Glasgow, Glasgow, United Kingdom\\
$^{61}$Oliver Lodge Laboratory, University of Liverpool, Liverpool, United Kingdom\\
% ...
$^{69}$Syracuse University, Syracuse, NY, United States\\
% ...
$^{87}$Department of Physics and Astronomy, Uppsala University, Uppsala, Sweden, associated to $^{60}$\\
% ...
}
\end{flushleft}




\end{document}